\documentclass[conference]{IEEEtran}
\IEEEoverridecommandlockouts
\usepackage{cite}
\usepackage{amsmath,amssymb,amsfonts}
\usepackage{algorithmic}
\usepackage{graphicx}
\usepackage{textcomp}
\usepackage{xcolor}
\usepackage{float}
\usepackage{wrapfig} 
\usepackage{makecell}
\usepackage{array}
\usepackage{xcolor} 

\def\BibTeX{{\rm B\kern-.05em{\sc i\kern-.025em b}\kern-.08em
    T\kern-.1667em\lower.7ex\hbox{E}\kern-.125emX}}
\begin{document}

\title{Programmable Governance \\for Group-Controlled \\Decentralized Identifiers
}

\author{\IEEEauthorblockN{Carlo Segat, Sandro Rodriguez Garzon, Axel Küpper}
\IEEEauthorblockA{\textit{Service-centric Networking} \\
\textit{Technische Universität Berlin / T-Labs}\\
Berlin, Germany \\
\{carlo.segat\}\textbar\{sandro.rodriguezgarzon\}\textbar\{axel.kuepper\}@tu-berlin.de}}

\maketitle

\begin{abstract}
Self-Sovereign Identity (SSI) is a paradigm for digital identity management that offers unique privacy advantages. A key technology in SSI is Decentralized Identifiers (DIDs) and their associated metadata, DID Documents (DDOs). DDOs contain crucial verification material such as the public keys of the entity identified by the DID (i.e., the DID subject) and are often anchored on a distributed ledger to ensure security and availability.

Long-lived DIDs need to support updates (e.g., key rotation). Ideally, only the DID subject should authorize DDO updates. However, in practice, update capabilities may be shared or delegated. While the DID specification acknowledges such scenarios, it does not define how updates should be authorized when multiple entities jointly control a DID (i.e., group control).

This article examines the implementation of an on-chain, trustless mechanism enabling DID controllers under group control to program their governance rules. The main research question is the following: Can a technical mechanism be developed to orchestrate on-chain group control of a DDO in a ledger-agnostic and adaptable manner?

\begin{IEEEkeywords}
SSI, DID, Governance, Smart Contract, Access Control
\end{IEEEkeywords}
\end{abstract}

\section{Introduction} 

Group control of ledger-anchored decentralized identifiers (DIDs) can be understood as a governance problem. In a nutshell, having control of a DID means having the authorization to update the associated DID Document (DDO), which can be understood as a container for key material and other metadata. With only one controller, this is a straightforward access-control problem, often tackled by the existing did:methods implementations by leveraging the built-in access control mechanisms of the underlying verifiable data registry (VDR). On the other hand, under group control, multiple entities share control of the DID. Hence, it becomes a multi-stakeholder access control problem, without an equally straightforward solution. This problem was already identified and analyzed in \textit{Governance of Ledger-Anchored Decentralized Identifiers} \cite{sandro_paper}, where the lack of comprehensive solutions both in the literature and in the concrete did:methods implementations was highlighted. This article addresses the same problem by presenting the design and implementation of a technical solution. The main research question can be understood as follows: is it possible to implement a flexible technical mechanism for managing on-chain group control of a DDO? Flexibility refers to having configurable controller authorization and coordination strategies. Another complementary research question is whether this can be done in a ledger-independent way.
 
DIDs are a foundational technology associated with Self-Sovereign Identity (SSI), a paradigm for privacy-preserving digital identity management. At a high level, SSI differs from centralized Identity Management Systems (IdMS) and federated ones by making the identity life-cycle independent from third parties. For a deeper discussion of SSI and the characteristics of did:methods, refer to \cite{ssi_the_book, awid_paper, did_core_w3c}. DIDs are unique, self-issued identifiers. Each DID is resolvable to a DID Document (DDO), a piece of metadata containing cryptographic material such as public keys and ZKP public commitment material. The decision of where to store the DDO, to make it persistent and retrievable, is a crucial architectural aspect. The W3C DID Specification refers to this storage layer as the VDR. Although centralized databases can realize the VDR, using a distributed ledger (DL) provides benefits in terms of data integrity, availability, auditability, and trust \cite{paper_motivating_use_of_dlt_as_vdr}. Another cornerstone of SSI is Verifiable Credentials (VCs). Trusted issuers issue VCs, containing cryptographically verifiable claims, bound to a
DID identified subject. VCs decouple the subject's claims from key material and improve flexibility compared to how this is handled in traditional PKI/x.509 systems. Moreover, VCs are a privacy-preserving technology as claims can be presented and verified without involving the issuer or another 3rd party \cite{w3c_vc_spec}.

In principle, the entity identified by a DID (i.e., the DID subject) is the sole controller (i.e,. the only authorized entity to update the DDO). In practice, however, control may need to be shared or delegated. For example, parents may wish to register their minor children's digital identities with a school; An employee temporarily becomes a deputy for his supervisor while the latter is on sick leave; or a legal entity participating
in a joint venture wants to sign a contract on the joint venture’s behalf. According to the DID specification, the DID subject can appoint other did-identified entities (i.e., deputies) to act on the DID subject's behalf. Technically, the DID subject inserts the deputies' public key material in the DDO to realize this. This enables deputies, e.g., to autonomously authenticate on behalf of the DID subject or to perform arbitrary updates to the DDO. 

Although this control-delegation mechanism is accounted for in the DID specification \cite{did_core_w3c}, the rules governing DDO updates under group control are not clarified. In turn, as revealed in \cite{sandro_paper}, most DID methods fall back to independent control, which means that each controller has full update rights. 

This paper presents a prototypical multi-stakeholder governance system that uses Smart Contracts to manage DL-anchored DIDs. When a DDO is anchored, its governance configuration must also be provided. Essentially, this configuration must answer those questions: how are the controllers identified? How is a controller authorized to act as such? What is the function that turns the controllers' decisions into a governance outcome? Crucially, this governance information is embedded in the DDO. When an entity wants to perform an update, they follow a well-defined on-chain governance process that consists of: proposal submission, coordination process, and proposal resolution. In the coordination phase, the controllers submit their decisions on an update proposal. A cost analysis of the prototype is also provided. This enables an assessment of system scalability and economic viability. Moreover, it provides a reference for comparison of the system with alternative solutions. To the best of our knowledge, the system is novel in 2  aspects. It supports \textbf{per-DID governance}, as each DID can be configured to be governed by different rules. It enables \textbf{governance evolution}, as the governance rules are part of the DDO and can therefore be updated with the same process for standard DDO content.

The paper is structured as follows. The related work section contains a summary of the existing research on the topic. Next, system design and prototype implementation are discussed. The evaluation section analyzes the costs and scalability properties of the proposed solution. Finally, in the conclusion, the contribution is summarized, and future work is outlined.

\section{Related Work}

The W3C specification mandates that for DIDs where the subject and controller are the same, the sole entity authorized to update the DDO is the subject. However, governance rules are not given when there is more than one controller. Two general scenarios are defined. In \textbf{Individual Control}, each controller has full privileges independently of the others. Whereas in \textbf{Multiple Control}, the controllers must \textit{"act together in some fashion"} \cite{did_core_w3c} to authorize changes. While individual control is readily understood, multiple control leaves room for interpretation, as coordination is prescribed without establishing clear boundaries. This flexibility allows for customization but also raises interoperability and security challenges, leaving implementers to develop their own authorization mechanisms.

Individual control does not enable fine-grained access control. As a consequence, as highlighted in prior research, this model is weak against governance attacks. In \cite{Mazzocca.2025}, the authors identify 2 attacks stemming from this limitation:
\begin{itemize}
  \item \textbf{Coup Attack} where a malicious controller removes all other controllers, monopolizing control over the DID.
  \item \textbf{Impersonation Attack} where a malicious controller replaces the legitimate DID subject’s public key with one under their control, locking out the original owner while gaining the ability to use the DID.
\end{itemize}
More authors address the issue, for example, \cite{vulnerability_did_doc_update} proposes to associate a delegation policy with each non-subject did controller. In this way, the capabilities of the controllers can be controlled. An "invocation history" feature is also proposed; every action performed by a controller is logged and can be monitored to detect misuse. 
Finally, in \cite{modid}, the authors propose a hierarchical controller structure where an N-of-M voting process is carried out before an update to a DDO is approved. The proposed system also includes a credential management part where VCs are stored on IPFS. Unfortunately, the focus is on key recovery, and the broader issue of DDO updates under multiple controllers is not addressed. Moreover, the system couples DDO controller management with credential management (i.e., who can upload VCs to IPFS and present them) without a justification.

Although the W3C DID Specification does not address how group control should be realized, some DID methods support this feature to various degrees of functionality. As noted in \cite{sandro_paper}, methods like \texttt{did:algo} \cite{didalgo}, \texttt{did:sol} \cite{didsol}, and \texttt{did:ebsi} \cite{didebsi} support simple individual control. In contrast, methods such as \texttt{did:factom} \cite{didfactom} and \texttt{did:iota} \cite{didiota} support group control. \texttt{did:iota}v2 recently introduced a multi-group control mechanism that aligns with the approach proposed in this paper. Controllers have weighted voting power, and updates follow a proposal-and-vote process.
One key problem is that the list of controllers is not stored in the DDO itself. To implement group control \textit{did:iota} leverages the IOTA multi-signature accounts. The DDO is owned by a single multi-signature account, and the controllers jointly use their private keys to instrument this account. Consequently, the ACL list is not directly captured within the DDO but is implicitly defined by the key provided as input when creating the multi-signature account. Nonetheless, there are some limitations: 
\begin{itemize}
  \item \textbf{Authorization Rigidity} as authentication relies solely on an access control list (ACL), with no support for alternative credential-based authorization (e.g., Verifiable Credentials). 
  \item \textbf{Governance Inflexibility} as threshold-based voting is the only supported coordination strategy, and customizations to voting are not supported. 
  \item  \textbf{VDR Dependency} as the system is tightly bound to the IOTA’s ledger, limiting portability. 
\end{itemize}
Similarly to \textit{did:iota}, most DID methods rely on implicit VDR-specific mechanisms instead of including the DID controllers in the DDO itself. This severely limits interoperability between did:methods and the transfer of DDOs between different VDRs.

ERC-1056 \cite{erc_1056} specifies how Ethereum accounts can be used as DIDs. It includes features such as ownership transfer and time-bound delegations. It enables every Ethereum account to be used as a DID without submitting a prior transaction, saving on cost. Another cost-saving aspect is how DDO updates are managed. Instead of saving updates as part of the SC state, the Ethereum event mechanism is used. In other words, an update to a DDO simply triggers the emission of an event. In turn, to reconstruct a DDO, one must collect all the emitted events relevant to it. 
In the context of SSI, ERC-725 \cite{erc_725} is an account abstraction standard where instead of EOAs, user accounts are SCs \cite{eth_account_abstraction}. Those accounts can be owned by an EOA or another SC. While group control could be realized by assigning ownership to a multi-signature account, ERC-725 is not meant to address governance issues as it focuses on low-level aspects of storage and transaction execution. 
SP-6-KeyManager (a successor of ERC-734) complements ERC-725 with key management features \cite{sp_6_key_manager}. It is a standard on the Lusko blockchain (an independent EVM-compatible blockchain \cite{lusko}). SP-6-KeyManager defines a SC that enforces access control on ERC-725 accounts (corresponding to PEP in the AAA framework). It supports multiple controllers (EOAs or SCs) and granular permissions (e.g., authorizing transfers, ownership changes, or specific function calls). Both keys and permission information are stored on the ERC-725 account and it allows the specifications of what keys a controller has the right to modify. However, while offering sophisticated access control, it still does not address group governance beyond individual permissions.



The AAA Authorization Framework \cite{rfc_aaa_policy_stuff} and its implementation XACML (eXtensible Access Control Markup Language) represent an important effort in standardizing access control. It defines roles and flows, establishing a clear mechanism to evaluate authorization requests to protected resources. The Policy Administration Point (PAP) is responsible for creating and managing access authorization policies. The Policy Decision Point (PDP) evaluates access requests against these policies to determine whether access should be granted or denied. The Policy Enforcement Point (PEP) acts as the gatekeeper, intercepting user requests, consulting the PDP for a decision, and enforcing the resulting access control decision. Additionally, the Policy Retrieval Point (PRP) serves as the storage repository for XACML policies, ensuring policies are available for retrieval. Given its prominence in the literature, in Section \ref{section:design} section a discussion on how components of this system map to the AAA Authorization Framework roles is included.

\section{System Design} 
\label{section:design}
\begin{figure}[t!]
\centerline{\includegraphics[scale=0.0825]{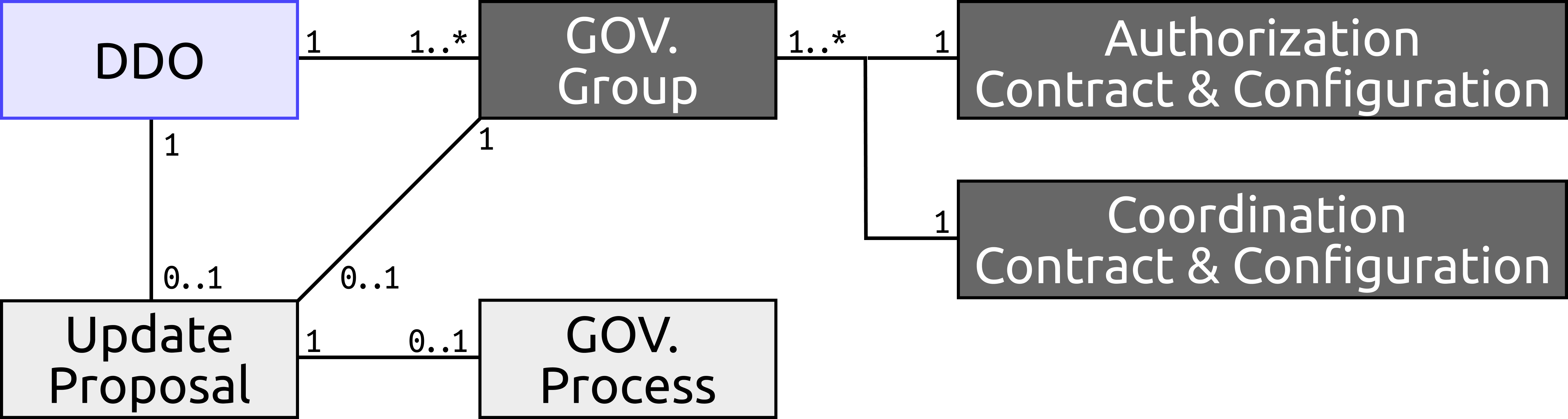}}
    \caption{Entity-relationship model of the proposed system. UML cardinality conventions are used.}
    \label{fig:arch_diagram}
\end{figure}

Figure \ref{fig:arch_diagram} shows the entity-relationship model of the system. The central element is the DDO. The DDO can be associated with at most one update proposal. If more proposals were active simultaneously, the first to successfully resolve would invalidate all the pending ones, as they would then be referring to an outdated version of the DDO. Although support for parallel proposals is achievable, the system is deliberately kept simple in this regard. However, in the implementation, an update proposal from a higher-privilege controller overrides and rejects an ongoing proposal from a lower-privilege one. This choice is important as it mitigates certain governance attacks where a malicious proposal is submitted to deny other controllers from submitting proposals. Crucially, an update proposal can include DDO content changes (e.g., key rotation, service-point additions, etc) as well as modifications to the rules governing the DDO. Each active proposal corresponds to a single governance process, representing the ongoing coordination process. 
The governance groups (GGs) associated with a DDO encode its governance rules. A GG is akin to a container for the following governance settings. 1) The address of an authorization contract and its configurations. 2) The address of a coordination contract and its configurations. Authorization and coordination contracts can be reused across different GGs. Finally, a DDO can be associated with more than one GG. Having multiple groups enables fine-grained, discretionary access control; a controller with the highest privilege can create a lower-privilege GG. Participation in the governance process is limited to the members of the GG from where the update proposal originated, hence the cardinality arrow between Update Proposal and Gov. Group. 

From the perspective of the controllers, two types of actions are subject to authorization: submitting an update proposal and a coordination decision. As said earlier, a key requirement is supporting flexible authorization rules. Current solutions are limited to static ACLs, but this design advances this by allowing users to establish their own authorization rules. The system components implementing those rules must adhere to well-defined interfaces, allowing other components to delegate authorization decisions by invoking them. To demonstrate this flexibility, alongside ACL authorization, two credential-based authorization schemes are considered.  
\begin{itemize}
    \item \textbf{Opaque bearer tokens} are simple tokens consisting of a nonce and a valid issuer signature.
    \item \textbf{VCs} are more sophisticated credentials that provide cryptographically verifiable claims about a subject and are non-transferable (bound to a holder's public key). This flexibility is useful in many ways. For example, it enables Role-Based Access Control (RBAC), where trusted issuers assign VCs that encode roles (e.g., admin, contributor) to controllers. 
\end{itemize}
Technically, also authorization of DDO anchoring needs to be discussed, but since the focus is on DDO update governance, it is deemed out of the scope of this paper.

Similarly to authorization rules, coordination rules are also configurable. This adaptability is useful because coordination requirements may vary depending on both the DDO's purpose and the organizational structure behind it. For example, adding a new board member to a company might demand unanimous consent, while a simple majority suffices for a routine key rotation. As the following two examples illustrate, coordination can take forms different than voting. 
\begin{itemize}
    \item \textbf{Hierarchical Approval} where an employee's update proposal requires their manager's approval, who in turn needs the higher-level manager's approval.
    2) \textbf{Preference Reconciliation} where the controllers submit a set of values representing a certain preferred system state; at resolution, the system tries to reconcile those values and, if successful, the agreed system state is updated to be the common denominator. 
\end{itemize} 

Furthermore, two additional governance aspects were considered: coordination execution, and time. Coordination execution refers to whether the coordination process occurs entirely on-chain or involves off-chain steps. It should be noted that proposal submission and resolution always happen entirely on-chain, but coordination decisions such as votes can be aggregated off-chain and submitted with a single transaction. Finally, each governance method can be time-limited, i.e., programmed to resolve after a certain amount of time has elapsed. 

To reiterate, the system is designed to be extensible in that the users can program any specialized authorization or coordination logic their particular scenario might require.

As shown in Table \ref{tab:aaa_mapping}, the roles defined in the AAA Authorization Framework can be broadly mapped to the components of the proposed system. However, the alignment is not exact, reflecting the difference between AAA's centralized authorization approach and the decentralized one taken here. 
One example is the PAP. At the beginning, as the governance rules (i.e., policies) are included in the DDO when it is anchored, the entity anchoring the DDO corresponds to the PAP. Subsequently, any controller can take this role as any of them can propose an update, potentially containing new governance rules.
\begin{table}
    \centering
    \begin{tabular}{|p{0.4cm}|p{3.5cm}|p{3.1cm}|}
        \hline
        \textbf{AAA} & \textbf{Purpose} & \textbf{This System}\\  \hline
        PEP & Intercepts request; consults PDP for decision; enforces decision & Router (proposal creation); coordination contract (coordination decision).\\  \hline
        PDP & Evaluates request against policies & Authorization contract\\  \hline
        PRP & Stores and makes policies available & DDO, authorization contract\\  \hline
        PAP & manages access policies & Controller that anchors; all controllers \\  \hline
    \end{tabular}
    \vspace{0.1cm}
    \caption{Comparison of the main elements in the AAA Authorization Framework against the elements of the system presented in this paper.}
    \label{tab:aaa_mapping}
\end{table}

\section{Implementation}
For the implementation, Solidity \cite{solidity_language} was chosen because of its mature tooling and ecosystem \cite{thesis_comparison_sc_techs}. Moreover, Solidity is supported on EVM-compatible ledgers other than Ethereum (e.g., Polygon \cite{polygon_white_paper} and BNB Smart Chain \cite{bnb_chain}). Importantly, Solidity is also supported by private DL technologies (e.g., Hyperledger Besu \cite{besu} and Fabric \cite{fabric_supports_evm_scs}). Development was carried out using Hardhat \cite{hardhat}. For simplicity, the prototype was not built to use a standard DID method such as did:ethr \cite{did_ehtr} or other methods based on ERC-1056 \cite{erc_1056}. Since the focus is on group control, it uses a simple DID registry, storing a mapping between ETH addresses (considered as DIDs) and DDOs (simple structs holding public keys and other attributes). For the same reason, access-control considerations such as DDO retrieval permissions are also out of scope. The prototype implements 3 voting strategies. 
\begin{itemize}
    \item for \textbf{n-of-m} approval requires a fixed number of votes. For governance configurations that don't statically encode the controllers in an ACL, the $m$ parameter is to be understood as a turnout threshold. 
    \item for \textbf{turnout-sensitive}, the approval threshold is adjusted based on the number of submitted votes.
    \item for \textbf{weighted}, each controller is associated with a weight, and approval requires the sum of the weights of the submitted votes to be above a threshold. 
\end{itemize} 
As discussed in Section \ref{section:design}, the implemented authorization modes are ACL, opaque toekn, and VC.
Since SCs lack support for time-based programming paradigms, time-limiting a governance process requires using external off-chain services. To this end, a simple Cron Scheduler contract was implemented. Its "schedule" method is invoked by a coordination SC when a time-limited proposal is initialized. The method emits an event containing the address of the coordination SC and the duration of the governance process. An external, off-chain service must monitor the emitted event and invoke (after the duration has expired) the coordination contract that resolves the proposal. This mechanism is essentially how third-party SC scheduling services, such as Chain Link Automation, \cite{chainlink_automation} work. Finally, by emitting Solidity events after each operation, the system governance history is preserved and auditability is enabled. The source code of the implementation can be found here: https://github.com/CarloSegat/Programmable-Governance-for-Group-Controlled-DIDs.

Table \ref{tab:gov_combos_table} offers a summary of the four configurable governance aspects (controller authorization, coordination strategy, coordination execution, time) and their considered practical realizations. The implemented VC Authorization (SC) can be configured with a set of keys and corresponding values (defined in the authorization configuration). When evaluating a request, it checks that the presented VC contains these keys with matching values. The implementation is limited to strict equality checks. 
\begin{table}
    \centering
    \begin{tabular}{|m{2.2cm}|m{1.6cm}|m{1.9cm}|m{1.2cm}|}
        \hline
        \textbf{Coordination Type} & \textbf{Authorization} & \textbf{Coordination Execution} & \textbf{Time} \\
        \hline
        turnout-sensitive & ACL & On-Chain & Unlimited \\
        n-of-m & Token & & \\
        weighted & VC & Off-Chain & Limited \\
        \hline
    \end{tabular}
    \vspace{0.1cm}
    \caption{Column header represents governance aspects. Column values are concrete embodiments realized by the implementation.}
    \label{tab:gov_combos_table}
\end{table}

Regarding fine-grained, discretionary access control, each GG has an Edit Right Level that limits what an update proposal originating from this group can attempt to change. In descending order of privilege, the Edit Right Levels are the following:
\begin{itemize}
    \item \textbf{All} applies no restrictions, even other GGs can be modified.
    \item \textbf{DelegatesCreation} has the same rights as SelfGovernance, but additionally, new GGs with Edit Right of Document can be created. 
    \item \textbf{SelfGovernance} permits a controller to propose updates only to the DDO and its own GG.
    \item \textbf{Document} allows only updates to the DDO. 
\end{itemize} 
In the implementation, there is a coupling between the Registry SC and the Edit Right Levels. The Registry requires the Edit Rights Level details to determine whether a higher-privileged update proposal should override an ongoing one. These details also control whether a new GG should be appended or replace an existing one. In summary, as the prototype stands, custom edit right levels are not supported. Nonetheless, by modifying the Registry, this feature could be attained.

\begin{figure*}[t!]
\centerline{\includegraphics[width=1.99\columnwidth]{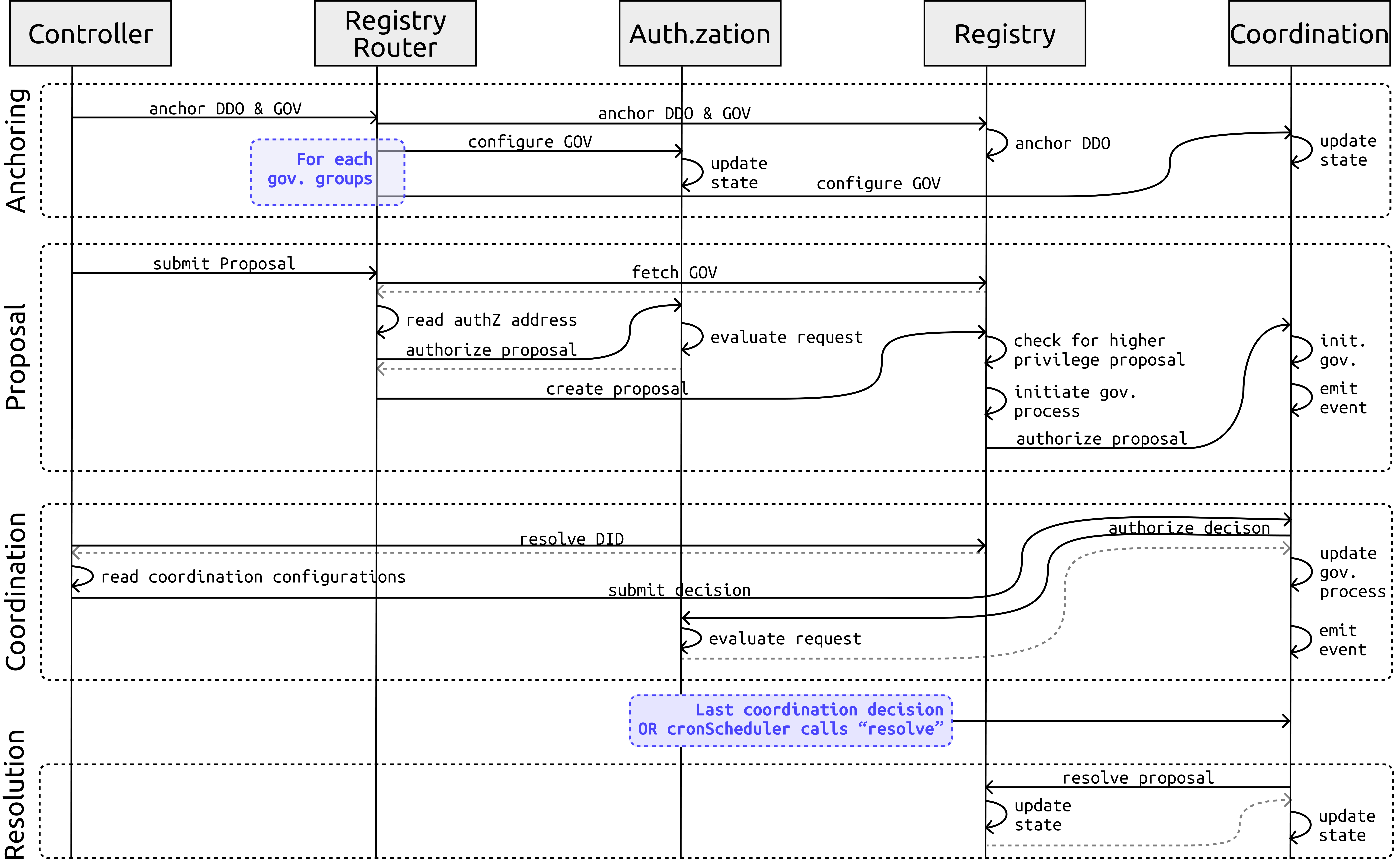}}
    \caption{High-level architecture diagram of the proposed system. The abbreviation "GOV" refers to governance.}
    \label{fig:seq}
\end{figure*}

The sequence diagram in Figure \ref{fig:seq} illustrates the DDO update process. Starting from anchoring, there are 4 main sections:
\begin{itemize}
    \item in the \textbf{Anchoring} phase, a controller submits a transaction to the Registry Router. This anchors the DDO and associated governance groups on the Registry contract. The Authorization and Coordination contracts are updated accordingly for each group.
    \item in the \textbf{Update Proposal} phase, upon receiving an update proposal from a controller, the Registry Router invokes the Authorization SC to authorize the proposal request. If authorization is granted, the Registry Router invokes the Registry SC, which updates its state and, in turn, calls the Coordination SC to initialize the coordination process. At the beginning of this process, the Registry Router must fetch the relevant governance configurations from the Registry to learn the address of the Authorization SC. 
    \item in the \textbf{Coordination Process} phase, each controller, after resolving the DID to learn the address of the Coordination SC, submits its decision to it. In the prototype, this is equivalent to voting, but, as stated in Section \ref{section:design}, different approaches are possible. Upon receiving a decision from a controller, the Coordination contract invokes the Authorization contract to authorize it. If successful, the Coordination contract updates its state to include the submitted decision.
    \item the \textbf{Resolution} phase is the final one where the coordination process is finalized. It is entered either after a decisive vote was cast or after the expiration of the governance process. In this phase, the Coordination contract invokes the Registry, which updates its state to reflect the outcome of the process. In other words, if the outcome was successful, the DDO and potentially its associated governance groups are updated.
\end{itemize}   

\section{Evaluation}
The system was evaluated by measuring static and dynamic gas costs. Static costs refer to SC deployments. Because static costs are paid once but dynamic costs are a repeating expense, it is important to evaluate them. The proposed solution is viable for production only if costs are sustainable. Costs are expressed in gas. A conversion to a monetary unit is not given because the gas price in Gwei is dynamic and changes with network conditions \cite{eth_gas_fee_how_works}. Nonetheless, after Ethereum Mainnet’s transition from Proof of Work (PoW) to Proof of Stake (PoS) consensus on 15 September 2022, gas prices have been falling. Before this date, the average was 47.0451 Gwei, but after, as of 5 July 2025, it decreased to 22.0822 Gwei. The Hardhat Node library \cite{hardhat} was used to simulate the workflows in Figure \ref{fig:seq} (anchoring, update proposal, coordination, and resolution). A script invokes the relevant SC methods and from the transaction receipt, the amount of gas used is extracted.

Table \ref{tab:deployment_costs} shows the deployment costs of the prototype SCs. Except for the relatively simple CronScheduler contract, deployment costs fall within the order of magnitude of $10^6$. The cost variations between contracts are not due to differences in the constructor logic (which is simple in all cases) but are directly related to the size of the deployed contract. Contract sizes in bytes are also shown. Importantly, they all fall below the 24K Bytes limit imposed by EIP-170 \cite{eip_170}.

\begin{table}
    \centering
    \begin{tabular}{|p{1.6cm}|p{2.6cm}|p{1.5cm}|p{1cm}|}
        \hline
        \textbf{System Part} & \textbf{Contract} & \textbf{Gas} & \textbf{Bytes} \\ \hline
    Core & Router & 1.2 × $1^6$ & 5478 \\
	& Registry & 3.1 × $1^6$ & 13819 \\
	& CronScheduler & 9.2 × $1^4$ & 177 \\ \hline
	Coordination & NOfM (nofm) & 2.4 × $1^6$ & 10682 \\
	& TurnoutSensitive (ts) & 2.5 × $1^6$ & 10985 \\
	& Weighted (w) & 2.4 × $1^6$ & 10878 \\ \hline
	Authorization & ACL (acl) & 1.1 × $1^6$ & 4839 \\
	& Token (tok) & 1.5 × $1^6$ & 6667 \\
	& VC (vc) & 1.8 × $1^6$ & 8069 \\ \hline
    \end{tabular}
    \vspace{0.1cm}
    \caption{Deployment costs of the system's smart contracts (gas) and deployed contract size (bytes).}
    \label{tab:deployment_costs}
\end{table}

\begin{wrapfigure}{l}{0.49\columnwidth} 
    \centering
    \includegraphics[height=0.5\columnwidth]{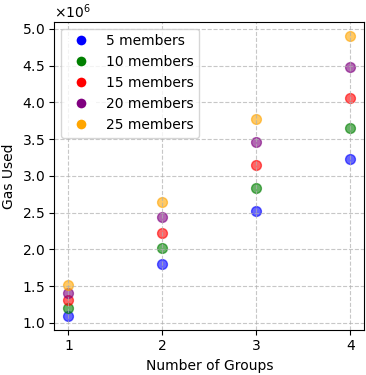}
    \caption{DDO Anchoring Cost vs. Number of Groups and Controllers}
    \label{fig:anchoring_overall}
\end{wrapfigure} 

Figure \ref{fig:anchoring_overall} shows the gas costs of DDO anchoring as it varies with the number of controller groups and controllers per group. The configurations of each governance group are the same for each data point (same number of members). The results show linear increases with both the number of governance groups and the number of members, matching expected iteration overhead. 

\begin{figure}[t]
    \centering
    \includegraphics[width=1\columnwidth]{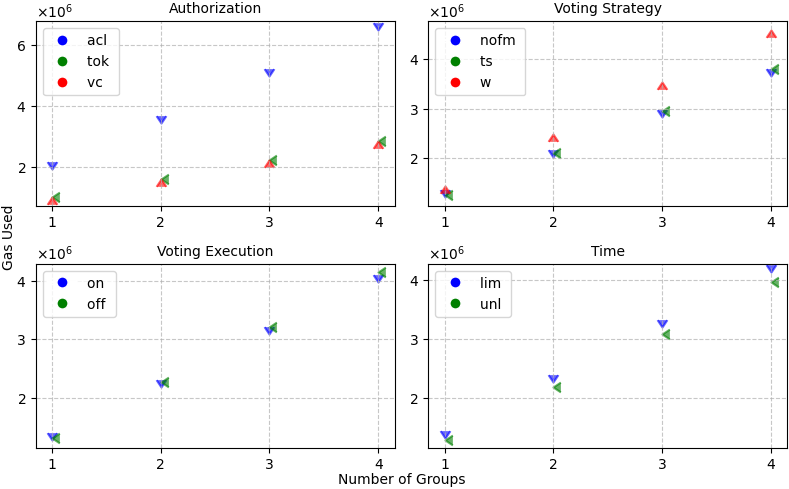}
    \caption{Anchoring cost (gas units) compared across governance dimensions.}
    \label{fig:anchoring_specific}
\end{figure}

Figure \ref{fig:anchoring_specific} compares the cost of anchoring between the 4 governance aspects. The labels in the legend correspond to those in Table \ref{tab:deployment_costs}. Each data point is obtained by averaging over the measurements that share the governance aspect of interest. One pattern that emerges is that ACL authorization and Weighted voting have slightly higher costs. This is due to storage overheads. Configurations that use ACL and Weighted need to store an array of weights (one for each controller). In others, weights are either not necessary or are found in the credentials (tokens and VCs). Each plot targets 1 governance aspect (i.e. authorization, coordination type, coordination execution, time). The time-limited configurations incur higher costs because they write the time-related settings and emit an additional event. 


\begin{figure}[t]
    \centering
    \includegraphics[width=1\columnwidth]{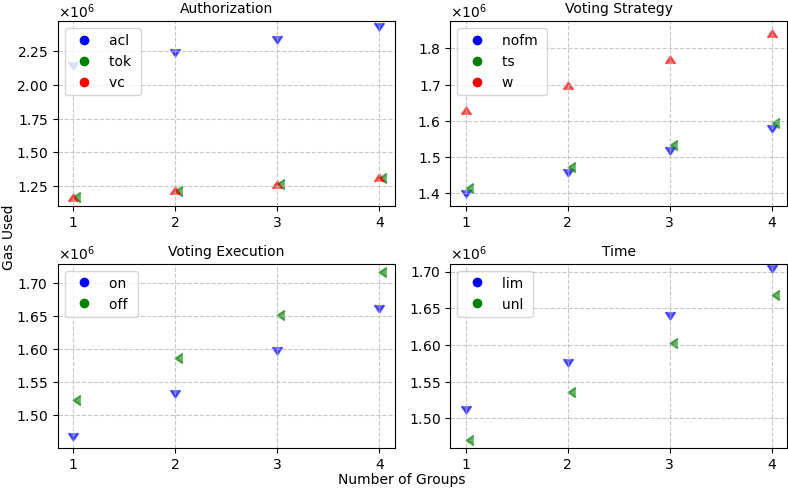}
    \caption{Proposal submission cost in gas units compared across governance dimensions.}
    \label{fig:proposal_specific}
\end{figure}

Figure \ref{fig:proposal_specific} shows the cost of submitting a proposal update. As seen before, it mirrors the direct proportionality between cost and the number of groups but the increase is more gradual due to fewer write operations compared to anchoring. This time, VC-authorization costs more than token authorization due to the additional holder signature check. As before, ACL-weighted configurations incur higher costs due to storage overhead. Moreover, a proposal that is time-limited or has off-chain voting requires additional setup operations, increasing costs.


\begin{figure}[t]
    \centering
    \includegraphics[width=1\columnwidth]{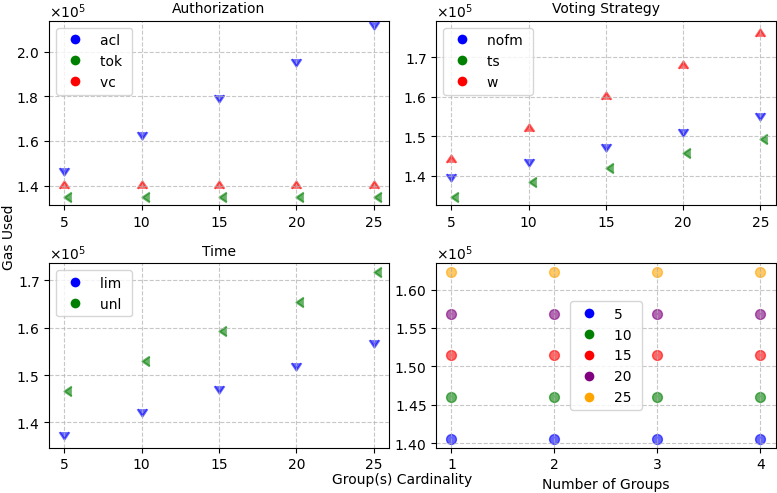}
    \caption{Vote submission cost in gas units compared across governance dimensions.}
    \label{fig:coordination_specific}
\end{figure}

\begin{figure}[t]
    \centering
    \includegraphics[width=1\columnwidth]{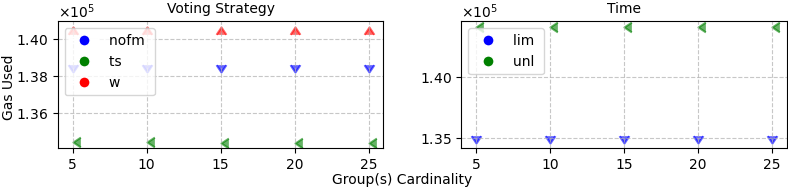}
    \caption{Vote submission cost in gas units compared across governance dimensions with the exclusion of configurations using ACL.}
    \label{fig:voting_no_acl}
\end{figure}

Figure \ref{fig:coordination_specific} shows the cost of single vote submission. As the bottom-right subplot indicates, this cost is independent of the number of governance groups, hence, the number of controllers is shown on the horizontal axis. For ACL, the cost increases with the number of controllers. This is due to the ACL contract logic iterating over the group members. Contrarily, for token and VC authorization, the cost is constant because they perform authorization by verifying signatures and do not require a static list. To show this, in Figure \ref{fig:voting_no_acl} the Voting Strategy and Time subplots are presented again, this time with the removal of ACL. Regarding voting strategies, weighted and N-of-M are the most expensive due to the execution of an iterative early-termination logic after each vote. The on/off-chain dimension is not relevant in this plot as submitting a single vote is only possible for on-chain governance. 



\begin{figure}[t]
    \centering
    \includegraphics[width=1\columnwidth]{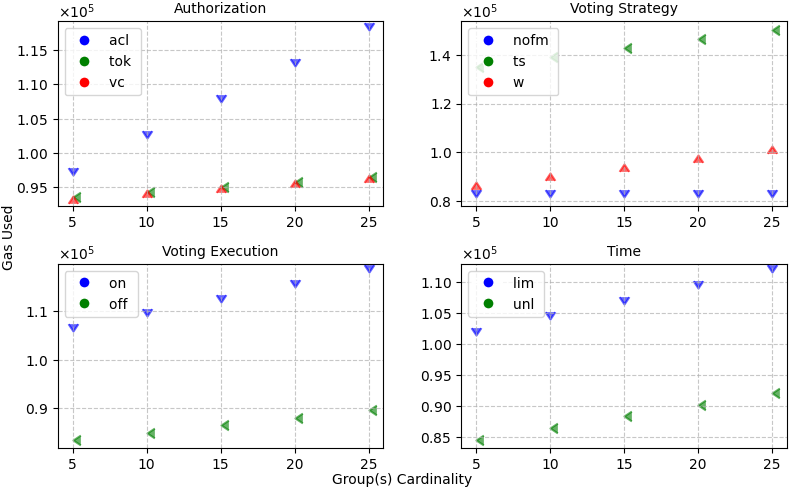}
    \caption{Resolution cost in gas units compared across governance dimensions.}
    \label{fig:resolution}
\end{figure}

Figure \ref{fig:resolution} shows the cost of resolving a proposal. While resolution for some coordination strategies can be triggered by a decision submitted by a controller (i.e., due to the early termination logic described previously), this is explicitly avoided. Data is generated by forcing manual resolution (after submitting a fixed number of decisions to avoid edge cases in the resolution logic). This isolates resolution from submitting a decision. As observed for vote submission, ACL authorization exhibits the steeper increase, with token and VC growing mildly with member count. Mild increases also occur for all voting strategies. Turnout-sensitive incurs in the highest costs because of its relatively more complex resolution logic.


\begin{figure}[t]
    \centering
    \includegraphics[width=1\columnwidth]{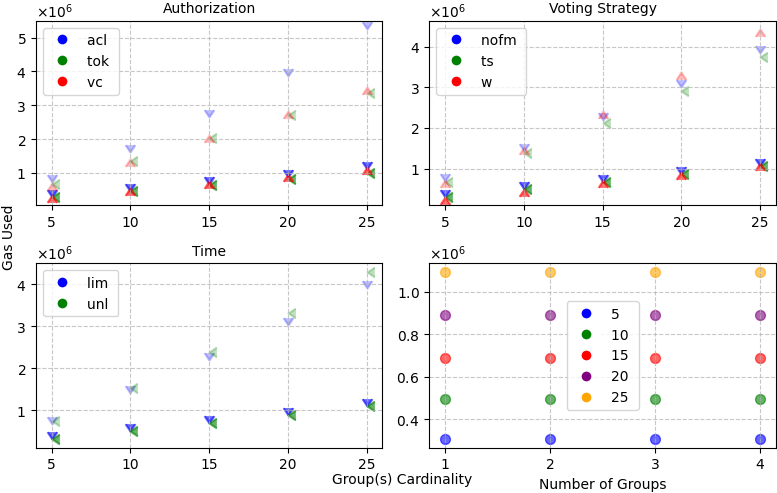}
    \caption{Off-chain voting cost in gas units compared across governance dimensions.}
    \label{fig:bulk_decison_submission}
\end{figure}

Figure \ref{fig:bulk_decison_submission} analyzes gas costs for off-chain voting, where participants' signatures are collected and submitted as a single transaction. In all cases, this approach (solid triangles) reduces costs significantly compared to submitting the same amount of individual on-chain decisions (transparent triangles). The bottom right of the figure shows the independence of off-chain coordination from the number of governance groups.


The data used to generate the figure is available here: https://github.com/CarloSegat/Programmable-Governance-for-Group-Controlled-DIDs.

\section{Conclusion}
This article presented the design of a programmable smart contract system for group control of ledger-based DIDs. Moreover, an open-source Solidity prototype was presented and published. The gas costs of deploying the system on a local Ethereum node, as well as the costs of anchoring a DDO,  submitting a proposal, carrying out the coordination process, and the resolution of a proposal are evaluated. It was shown that on-chain group control for ledger-anchored DDOs is feasible with a large degree of flexibility. Paramount to support governance evolution is the inclusion of governance configurations in the DDO itself, rather than storing them separately or relying on implicit mechanisms of the underlying DLT. The measurements show how credential-based authorization, compared to ACL, consumes less gas in recurring operations such as submitting an update proposal and voting on it. This occurs because ACL configurations incur additional memory accesses, proportional to the number of controllers. Nevertheless, this result relies on the assumption that the number of trusted issuers, encoded in the governance configuration, grows sublinearly compared to the number of controllers of an ACL. Because the developed system is compatible with all EVM-based ledgers, the provided gas cost analysis is useful to system architects evaluating the feasibility of basing an IdMS on such ledgers. Placing the information required to run governance in the DDO is a step forward in the realization of ledger-agnostic group control of DID. Yet, the system developed as part of this paper is not ledger agnostic, as it is coupled to Ethereum SC and account system. 
In summary, governing group-controlled, ledger-anchored DIDs presents significant complexity. This paper represents a first step towards systematically defining an overarching framework to address it. Future work should prioritize the development of a threat model to analyze the resiliency of such a system against both traditional security risks and governance vulnerabilities. Moreover, other aspects that wold advance the design and the system include: anchoring governance; cost sharing (e.g. now, only one controller bears the costs of anchoring and proposal submissions); integration with existing account solutions such as ERC-725, or use standardized did methods such as did:ethr; more sophisticated off-chain coordination beyond simple signature aggregation such as threshold signatures; Solidity optimizations, for example struct packing, bit packing and Bloom filters. 

\section{Acknowledgements}
This work was funded by the Federal Ministry of Education and Research (BMBF) in Germany under the grant number 16KIS2251 of the SUSTAINET\_guarDian project.

\bibliographystyle{IEEEtran}  
\bibliography{references} 

\end{document}